\documentclass[preprintnumbers,amsmath,amssymbm,prd]{revtex4}
\usepackage{epsfig}
\usepackage{graphicx}
\usepackage{amssymb}

\begin{document}
\title{A no-short scalar hair theorem for rotating Kerr black holes}
\author{Shahar Hod}
\affiliation{The Ruppin Academic Center, Emeq Hefer 40250, Israel}
\affiliation{ } \affiliation{The Hadassah Institute, Jerusalem
91010, Israel}
\date{\today}

\begin{abstract}
If a black hole has hair, how short can this hair be? A partial
answer to this intriguing question was recently provided by the
`no-short hair' theorem which asserts that the external fields of a
spherically-symmetric electrically neutral hairy black-hole
configuration must extend beyond the null circular geodesic which
characterizes the corresponding black-hole spacetime. One naturally
wonders whether the no-short hair inequality
$r_{\text{hair}}>r_{\text{null}}$ is a generic property of all
electrically neutral hairy black-hole spacetimes? In this paper we
provide evidence that the answer to this interesting question may be
positive. In particular, we prove that the recently discovered
cloudy Kerr black-hole spacetimes -- non-spherically symmetric
non-static black holes which support linearized massive scalar
fields in their exterior regions -- also respect this no-short hair
lower bound. Specifically, we {\it analytically} derive the lower
bound $r_{\text{field}}/r_+>r_+/r_-$ on the effective lengths of the
external bound-state massive scalar clouds (here $r_{\text{field}}$
is the peak location of the stationary bound-state scalar fields and
$r_{\pm}$ are the horizon radii of the black hole). Remarkably, this
lower bound is universal in the sense that it is independent of the
physical parameters (proper mass and angular harmonic indices) of
the exterior scalar fields. Our results suggest that the lower bound
$r_{\text{hair}}>r_{\text{null}}$ may be a general property of
asymptotically flat electrically neutral hairy black-hole
configurations.
\end{abstract}
\bigskip
\maketitle

\section {Introduction.}

The elegant uniqueness theorems \cite{Uni1,Uni2,Uni3,Uni4} have
established the fact that all stationary black-hole solutions of the
Einstein-vacuum equations are uniquely described by the Kerr
\cite{Kerr,Chan} spacetime metric. In his celebrated `no-hair'
conjecture, Wheeler \cite{Whee,Car} went even one step further by
suggesting that the Kerr spacetime is the only stationary black-hole
solution \cite{Noteeln} of the coupled Einstein-matter field
equations.

Matter fields in an asymptotically flat black-hole spacetime are
expected, according to the no-hair conjecture \cite{Whee,Car}, to be
scattered away to infinity or to be absorbed into the black hole.
Early analytical studies \cite{Chas,Hart,BekVec} of the coupled
Einstein-matter field equations have provided support for the
validity of this conjecture. In particular, these studies
\cite{Chas,Hart,BekVec} have shown that static scalar fields, static
spinor fields, and static massive vector fields cannot be supported
in the exterior spacetime regions of asymptotically flat regular
\cite{Notereg} black holes.

However, subsequent numerical studies of the non-linear
Einstein-matter field equations have revealed that other matter
models may be characterized by non-trivial (that is, non-Kerr like)
hairy black-hole solutions. In particular, the first convincing
counterexample to the no-hair conjecture was provided by the
`colored' hairy black-hole spacetimes \cite{BizCol}. These
non-trivial solutions of the coupled Einstein-Yang-Mills equations
describe non-vacuum black holes which support regular Yang-Mills
fields in their exterior regions.

The intriguing discovery of the Einstein-Yang-Mills hairy black-hole
solutions \cite{BizCol} resulted in an intense research effort to
understand the physical properties of these (and other
\cite{BizCol,Lavr,BizCham,Green,Stra,BiWa,EYMH,Volkov,BiCh,Lav1,Lav2,Bizw})
hairy black-hole spacetimes. These studies
\cite{BizCol,Lavr,BizCham,Green,Stra,BiWa,EYMH,Volkov,BiCh,Lav1,Lav2,Bizw}
have established the fact that various types of non-linear matter
models
\cite{BizCol,Lavr,BizCham,Green,Stra,BiWa,EYMH,Volkov,BiCh,Lav1,Lav2,Bizw},
when coupled to the Einstein field equations, may produce hairy
black-hole solutions that violate the original formulation
\cite{Whee,Car} of the no-hair conjecture \cite{Notevio}.

A generic feature of these hairy black-hole solutions
\cite{BizCol,Lavr,BizCham,Green,Stra,BiWa,EYMH,Volkov,BiCh,Lav1,Lav2,Bizw}
was revealed in \cite{Hod11}, where it was proved that if a static
spherically-symmetric black hole has hair, then this hair cannot be
short in the sense that the corresponding external matter fields
must extend beyond the null circular geodesic which characterizes
the black-hole spacetime \cite{Hod11,Hodaa,Notenun}:
\begin{equation}\label{Eq1}
r_{\text{field}}>r_{\text{null}}\  .
\end{equation}
It is important to emphasize the fact that this `{\it no-short
hair}' theorem was proved in \cite{Hod11} under the following three
assumptions:
\newline
(1) The hairy black-hole spacetime is static.
\newline
(2) The hairy black-hole spacetime is spherically-symmetric.
\newline
(3) The energy density $\rho\equiv T^t_t$ outside the black-hole
horizon approaches zero asymptotically faster than $r^{-4}$
\cite{Notec1}. In particular, the hairy black-hole spacetime is
assumed to be electrically neutral.

It is worth noting that, under the above-mentioned assumptions, the
no-short hair relation (\ref{Eq1}) is universal in the sense that it
is independent of the physical parameters of the external matter
fields \cite{Hod11,Hodaa,Notenun}. The no-short hair lower bound
(\ref{Eq1}) may therefore be regarded as a more modest alternative
(and possibly a more robust alternative) to the original
\cite{Whee,Car} no-hair conjecture.

One naturally wonders whether the no-short hair property (\ref{Eq1})
is a generic feature of {\it all} asymptotically flat electrically
neutral hairy black-hole spacetimes? The main goal of the present
paper is to test the validity of the no-short hair lower bound
(\ref{Eq1}) beyond the restricted regime of static
spherically-symmetric black-hole spacetimes. To that end, we shall
study analytically the physical properties of ({\it non}-static,
{\it non}-spherically symmetric) rotating Kerr black holes
\cite{Notec2} linearly coupled to stationary bound-state
configurations of massive scalar fields.

Before proceeding, it is worth noting that former analytical studies
\cite{Hodrc} have shown that stationary bound-state configurations
of massive scalar fields linearly coupled to near-extremal
\cite{Notene} Kerr black holes conform to the lower bound
(\ref{Eq1}). Moreover, {\it numerical} studies \cite{HerR} of this
non-static non-spherically symmetric physical system have provided
further compelling evidence that the composed
Kerr-black-hole-massive-scalar-field configurations respect the
lower bound (\ref{Eq1}). In the present paper we shall provide a
rigorous {\it analytical} proof for the validity of this no-short
hair property for generic \cite{Noteal}
Kerr-black-hole-massive-scalar-field configurations.

\section{Composed Kerr-black-hole-massive-scalar-field configurations.}

Recent analytical \cite{Hodrc} and numerical \cite{HerR} studies of
the coupled Einstein-Klein-Gordon field equations have revealed
that, due to the intriguing physical phenomenon of superradiant
scattering of bosonic (integer-spin) fields in Kerr black-hole
spacetimes \cite{Zel,PressTeu1}, these {\it rotating} black holes
can support stationary (that is, {\it non}-decaying) scalar field
configurations in their exterior spacetime regions.

These stationary bound-state regular field configurations mark the
boundary between stable and unstable resonances of the composed
Kerr-black-hole-massive-scalar-field system. In particular, for a
given value of the azimuthal harmonic index $m$, these orbiting
field configurations are characterized by azimuthal frequencies
$\omega_{\text{field}}$ which coincide with the threshold (critical)
frequency \cite{Noteunits}
\begin{equation}\label{Eq2}
\omega_{\text{field}}=\omega_{\text{c}}\equiv m\Omega_{\text{H}}\
\end{equation}
for superradiant amplification of bosonic fields in the rotating
Kerr black-hole spacetime. Here \cite{Chan,Kerr}
\begin{equation}\label{Eq3}
\Omega_{\text{H}}={{a}\over{r^2_++a^2}}\
\end{equation}
is the Kerr black-hole angular velocity, where $r_+$ and $a$
\cite{Notesmp} are the horizon-radius and angular momentum per unit
mass of the rotating Kerr black hole.

The resonance condition (\ref{Eq2}), which characterizes the
stationary scalar field configurations in the rotating Kerr
black-hole spacetime, guarantees that there is no net energy flux
into the black hole \cite{Hodrc,HerR,Dolh}. In addition, for a
scalar field of mass $\mu$, the mutual gravitational attraction
between the central black hole and the external massive field
provides a natural confinement mechanism which prevents the orbiting
scalar configuration from radiating its energy to infinity. In
particular, the external bound-state massive scalar configurations
are characterized by trapped field modes in the bounded frequency
regime \cite{Notedim} [see Eq. (\ref{Eq11}) below]
\begin{equation}\label{Eq4}
0<\omega^2_{\text{field}}<\mu^2\  .
\end{equation}

Before proceeding, we would like to emphasize that the bound-state
Kerr-black-hole-massive-scalar-field configurations that we shall
study in this paper should not be regarded as genuine hairy
black-hole spacetimes. In particular, we shall treat the external
stationary scalar field configurations at the {\it linear} level.
Hence, throughout the paper we shall use the term `scalar clouds' to
describe these linearized bound-state field configurations
\cite{Notehe}. While the fully non-linear Einstein-scalar field
equations can only be studied {\it numerically} \cite{HerR}, below
we shall explicitly demonstrate that the physical properties of the
linearly coupled Kerr-black-hole-massive-scalar-field configurations
can be studied {\it analytically} \cite{Noterg}.

\section{Description of the system.}

We shall explore the physical properties of a rotating Kerr black
hole of mass $M$ and angular-momentum $J\equiv Ma$ which is linearly
coupled to a scalar field $\Psi$ of mass $\mu$ \cite{Notedim}. In
terms of the Boyer-Lindquist coordinates $(t,r,\theta,\phi)$, the
black-hole spacetime metric is described by the line element
\cite{Chan,Kerr}
\begin{eqnarray}\label{Eq5}
ds^2=-{{\Delta}\over{\rho^2}}(dt-a\sin^2\theta
d\phi)^2+{{\rho^2}\over{\Delta}}dr^2+\rho^2
d\theta^2+{{\sin^2\theta}\over{\rho^2}}\big[a
dt-(r^2+a^2)d\phi\big]^2\  ,
\end{eqnarray}
where $\Delta\equiv r^2-2Mr+a^2$ and $\rho^2\equiv
r^2+a^2\cos^2\theta$. The zeros of $\Delta$ determine the black-hole
horizon radii:
\begin{equation}\label{Eq6}
r_{\pm}=M\pm\sqrt{M^2-a^2}\  .
\end{equation}

The Klein-Gordon field equation
\begin{equation}\label{Eq7}
(\nabla^{\nu}\nabla_{\nu}-\mu^2)\Psi=0\
\end{equation}
determines the dynamics of the linearized massive scalar field
$\Psi$ in the curved black-hole spacetime. It is convenient to write
the scalar eigenfunction $\Psi$ in the form \cite{Noteanz}
\begin{equation}\label{Eq8}
\Psi(t,r,\theta,\phi)=\sum_{l,m}e^{im\phi}{S_{lm}}(\theta;m,a\sqrt{\mu^2-\omega^2_{\text{c}}})
{R_{lm}}(r;M,a,\mu,\omega_{\text{c}})e^{-i\omega_{\text{c}} t}\  ,
\end{equation}
in which case one finds that the Klein-Gordon wave equation
(\ref{Eq7}) can be expressed as a set of two coupled ordinary
differential equations: the first equation [see Eq. (\ref{Eq9})
below] determines the angular part $S_{lm}$ of the scalar
eigenfunction $\Psi$, whereas the second equation [see Eq.
(\ref{Eq10}) below] determines the radial part $R_{lm}$ of the
scalar eigenfunction $\Psi$.

The characteristic angular equation (also known as the spheroidal
wave equation) is given by \cite{Stro,Heun,Fiz1,Teuk,Abram,Hodasy}
\begin{eqnarray}\label{Eq9}
{1\over {\sin\theta}}{{d}\over{\theta}}\Big(\sin\theta {{d
S_{lm}}\over{d\theta}}\Big)
+\Big[K_{lm}+a^2(\mu^2-\omega^2_{\text{c}})\sin^2\theta-{{m^2}\over{\sin^2\theta}}\Big]S_{lm}=0\
.
\end{eqnarray}
The angular solutions ${S_{lm}}(\theta)$ \cite{Notesa} of
(\ref{Eq9}) are required to be regular at the two boundaries,
$\theta=0$ and $\theta=\pi$. These boundary conditions single out a
discrete family $\{K_{lm}\}$ of angular eigenvalues which
characterize the massive scalar fields (see \cite{Barma,Yang,Hodpp}
and references therein).

The radial Klein-Gordon equation (also known as the radial Teukolsky
equation) is given by \cite{Teuk,Stro}
\begin{equation}\label{Eq10}
\Delta{{d}
\over{dr}}\Big(\Delta{{dR_{lm}}\over{dr}}\Big)+\Big\{[\omega_{\text{c}}(r^2+a^2)-ma]^2
+\Delta[2ma\omega_{\text{c}}-\mu^2(r^2+a^2)-K_{lm}]\Big\}R_{lm}=0\ .
\end{equation}
Note that the radial equation (\ref{Eq10}) is coupled to the angular
equation (\ref{Eq9}) \cite{Notenam,Notebj}. The stationary
bound-state massive scalar clouds, which are supported in the Kerr
black-hole spacetime, are characterized by exponentially decaying
(bounded) radial eigenfunctions at spatial infinity
\cite{Hodrc,HerR,Ins2}:
\begin{equation}\label{Eq11}
R(r\to\infty)\sim e^{-\sqrt{\mu^2-\omega^2_{\text{c}}}r}\  .
\end{equation}
In addition, regular field configurations in the black-hole
spacetime are characterized by finite radial eigenfunctions. In
particular \cite{Notepp},
\begin{equation}\label{Eq12}
0\leq R(r=r_+)<\infty\  .
\end{equation}

\section{The effective radial potential of the composed Kerr-massive-scalar-field configurations.}

In order to analyze the spatial properties of the linearized
bound-state massive scalar configurations (the stationary scalar
clouds) in the Kerr black-hole spacetime, we shall first write the
radial Teukolsky equation (\ref{Eq10}) in the form of a
Schr\"odinger-like wave equation. Defining the new radial function
\begin{equation}\label{Eq13}
\psi=rR\  ,
\end{equation}
and using the new radial coordinate $y$ which is defined by the
relation \cite{Notemap}
\begin{equation}\label{Eq14}
dy={{r^2}\over{\Delta}}dr\  ,
\end{equation}
one can write the radial Teukolsky equation (\ref{Eq10}) in the
compact form
\begin{equation}\label{Eq15}
{{d^2\psi}\over{dy^2}}-V(y)\psi=0\  .
\end{equation}
The effective radial potential in the Schr\"odinger-like wave
equation (\ref{Eq15}) is given by
\begin{equation}\label{Eq16}
V=V(r;M,a,\mu,l,m)={{2\Delta}\over{r^6}}(Mr-a^2)+{{\Delta}\over{r^4}}
[K_{lm}-2ma\omega_{\text{c}}+\mu^2(r^2+a^2)]-{{1}\over{r^4}}[\omega_{\text{c}}(r^2+a^2)-ma]^2\
.
\end{equation}

In the next section we shall explore the near-horizon behavior of
the effective radial potential (\ref{Eq16}) and the corresponding
spatial properties of the associated radial eigenfunction $\psi$.

\section{The near-horizon behavior of the radial eigenfunctions.}

In the present section we shall analyze the near-horizon properties
of the radial eigenfunction $\psi$ which characterizes the
stationary bound-state resonances of the massive scalar fields in
the rotating Kerr black-hole spacetime. In particular, we shall
prove that $\psi$ is a positive \cite{Notepp}, increasing, and
convex function in the near-horizon $r-r_+\ll r_+-r_-$ region. To
that end, we shall first analyze the spatial behavior of the
effective radial potential (\ref{Eq16}) in the near-horizon region.

Substituting into (\ref{Eq16}) the resonant frequency (\ref{Eq2}) of
the stationary scalar field, one finds the near-horizon behavior
\begin{equation}\label{Eq17}
r^2_+V(x\to0)=F\tau\cdot x+O(x^2)\
\end{equation}
of the effective radial potential, where we have used here the
dimensionless variables
\begin{equation}\label{Eq18}
x\equiv {{r-r_+}\over{r_+}}\ \ \ \ ; \ \ \ \
\tau\equiv{{r_+-r_-}\over{r_+}}\  .
\end{equation}
The expansion coefficient in (\ref{Eq17}) is given by
\begin{equation}\label{Eq19}
F\equiv K_{lm}-{{2(ma)^2}\over{r^2_++a^2}}+\mu^2(r^2_++a^2)+\tau\ .
\end{equation}
Taking cognizance of the inequality (\ref{Eq4}) and using the lower
bound \cite{Barma,Notesi}
\begin{equation}\label{Eq20}
K_{lm}\geq m^2-a^2(\mu^2-\omega^2_c)\
\end{equation}
on the angular eigenvalues, one finds the characteristic inequality
\begin{equation}\label{Eq21}
F>m^2\cdot{{r^2_+}\over{r^2_++a^2}}+\tau>0\  .
\end{equation}
Taking cognizance of Eqs. (\ref{Eq17}) and (\ref{Eq21}), one deduces
that, in the near-horizon $x\ll\tau$ region, the radial potential
(\ref{Eq16}) takes the form of an effective potential barrier with
$V\geq0$.

Using Eqs. (\ref{Eq14}) and (\ref{Eq18}), one finds the relation
\begin{equation}\label{Eq22}
y={{r_+}\over{\tau}}\ln(x)+O(x)\
\end{equation}
in the near-horizon region
\begin{equation}\label{Eq23}
x\ll\tau\  .
\end{equation}
This relation can also be written in the form \cite{Noteyas}
\begin{equation}\label{Eq24}
x=e^{\tau y/r_+}[1+O(e^{\tau y/r_+})].
\end{equation}
Using Eqs. (\ref{Eq17}) and (\ref{Eq24}), one finds that, in the
near-horizon region (\ref{Eq23}), the Schr\"odinger-like radial
equation (\ref{Eq15}) can be written in the form
\begin{equation}\label{Eq25}
{{d^2\psi}\over{d\tilde y^2}}-{{4F}\over{\tau}}e^{2\tilde y}\psi=0\
,
\end{equation}
where
\begin{equation}\label{Eq26}
\tilde y\equiv {{\tau}\over{2r_+}}y\  .
\end{equation}

The near-horizon radial equation (\ref{Eq25}) can be solved
analytically. In particular, the solution of (\ref{Eq25}) which
respects the boundary condition (\ref{Eq12}) can be expressed in
terms of the modified Bessel function of the first kind
\cite{Noteab1,Notesk}:
\begin{equation}\label{Eq27}
\psi(y)=I_0\Big(2\sqrt{{{F}\over{\tau}}}e^{\tau y/2r_+}\Big)\  .
\end{equation}
Using the well-known properties of the modified Bessel function
$I_0$ \cite{Abram}, one deduces from (\ref{Eq27}) that, in the
near-horizon $x\ll\tau$ region, the radial eigenfunction $\psi$ is a
positive, increasing, and convex function. That is,
\begin{equation}\label{Eq28}
\{\psi>0\ \ \ ;\ \ \ {{d\psi}\over{dy}}>0\ \ \ ;\ \ \
{{d^2\psi}\over{dy^2}}>0\}\ \ \ \ \text{for}\ \ \ \ 0<x\ll\tau\ .
\end{equation}

Taking cognizance of the near-horizon behavior (\ref{Eq28})
\cite{Noteep1} and the far-region asymptotic behavior (\ref{Eq11})
\cite{Noteep2} of the radial eigenfunction $\psi$, one arrives at
the important conclusion that this function, which characterizes the
stationary bound-state scalar configurations in the Kerr black-hole
spacetime, must have (at least) one maximum point,
$x=x_{\text{max}}$, in the black-hole exterior region.

\section{A lower bound on the effective lengths of the stationary
bound-state Kerr scalar clouds.}

We have seen that the radial eigenfunction $\psi$, which
characterizes the stationary bound-state configurations of the
massive scalar fields in the rotating Kerr black-hole spacetime, is
a {\it non}-monotonic function. In particular, we have proved that
$\psi$ must have (at least) one maximum point outside the black-hole
horizon. In the present section we shall obtain a generic lower
bound on the peak location, $r_{\text{max}}$, of these bound-state
massive scalar configurations.

We first point out that the radial function $\psi$ must have an
inflection point, $r=r_0$ \cite{Notebr}, somewhere in the interval
$(r_+,r_{\text{max}})$. That is,
\begin{equation}\label{Eq29}
r_+<r_0<r_{\text{max}}\  .
\end{equation}
Taking cognizance of the radial equation (\ref{Eq15}), one deduces
that this inflection point [with ${{d^2\psi}/{dy^2}}=0$ at
$y_0=y_0(r_0)$] is a turning point of the effective radial potential
(\ref{Eq16}). That is,
\begin{equation}\label{Eq30}
V(r=r_0)=0\  .
\end{equation}
We shall now derive a lower bound on the radial location of this
inflection point.

Substituting into (\ref{Eq16}) the resonant frequency (\ref{Eq2}) of
the stationary scalar field, and using the characteristic
inequalities (\ref{Eq4}) and (\ref{Eq20}), on finds the lower bound
\begin{equation}\label{Eq31}
V(r)>m^2\cdot
{{(r-r_+)(r^2_+-rr_-)}\over{r^3(r^2_++a^2)}}+{{2\Delta}\over{r^6}}(Mr-a^2)\
\end{equation}
on the effective radial potential. Furthermore, using the inequality
$Mr-a^2\geq Mr_+-a^2=r_+(M-r_-)\geq0$, one obtains from (\ref{Eq31})
the characteristic inequality
\begin{equation}\label{Eq32}
V(r)>m^2\cdot {{r-r_+}\over{r^3(r^2_++a^2)}}\times(r^2_+-rr_-)\ .
\end{equation}
Taking cognizance of Eqs. (\ref{Eq30}) and (\ref{Eq32}), one finds
the lower bound
\begin{equation}\label{Eq33}
r_0>{{r^2_+}\over{r_-}}
\end{equation}
on the radial location of the inflection point $r=r_0$ which
characterizes the radial eigenfunction $\psi$.

Finally, taking cognizance of the inequalities (\ref{Eq29}) and
(\ref{Eq33}), one obtains the lower bound
\begin{equation}\label{Eq34}
r_{\text{max}}>{{r^2_+}\over{r_-}}
\end{equation}
on the peak location $r_{\text{max}}$ of the radial eigenfunction
$\psi$ which characterizes the stationary bound-state massive scalar
configurations in the rotating Kerr black-hole spacetime.
Remarkably, this lower bound is universal in the sense that it is
independent of the physical parameters (proper mass and angular
harmonic indices) of the external massive scalar fields.

\section{Stationary bound-state Kerr scalar clouds and null circular geodesics.}

In the present section we shall show that the composed
Kerr-black-hole-massive-scalar-field configurations respect the
no-short hair lower bound (\ref{Eq1}) \cite{Noteepm}. In particular,
we shall prove that the peak location $r_{\text{max}}$ of the radial
eigenfunction $\psi$, which characterizes the external bound-state
scalar clouds, is located beyond the equatorial null circular
geodesic of the corresponding black-hole spacetime.

The equatorial null circular geodesics of the rotating Kerr
black-hole spacetimes are characterized by the relation \cite{Chan}
\begin{equation}\label{Eq35}
r_{\text{null}}=2M\{1+\cos[{2\over 3}\cos^{-1}(-a/M)]\}\  .
\end{equation}
Taking cognizance of (\ref{Eq6}) and (\ref{Eq35}), one finds that
the ratio $r_{\text{null}}r_-/r^2_+$ is a monotonic increasing
function of the black-hole rotation parameter $a$. Specifically,
this dimensionless ratio increases from $0$ to $1$ as $a/M$
increases from $0$ to $1$. That is,
\begin{equation}\label{Eq36}
{{r_{\text{null}}r_-}\over{r^2_+}}\leq1\  .
\end{equation}
On the other hand, from Eq. (\ref{Eq34}) one finds the
characteristic inequality \cite{Notels}
\begin{equation}\label{Eq37}
{{r_{\text{max}}r_-}\over{r^2_+}}>1
\end{equation}
for the external bound-state scalar clouds. Taking cognizance of
(\ref{Eq36}) and (\ref{Eq37}), one concludes that the stationary
bound-state scalar configurations of the rotating Kerr black-hole
spacetime are characterized by the relation
\begin{equation}\label{Eq38}
r_{\text{max}}>r_{\text{null}}\  .
\end{equation}

\section{Summary.}

The `no-short hair' theorem \cite{Hod11} asserts that the external
matter fields of a static spherically-symmetric electrically neutral
hairy black-hole configuration must extend beyond the null circular
geodesic which characterizes the corresponding black-hole spacetime.

The main goal of the present paper was to test the validity of the
no-short hair lower bound (\ref{Eq1}) beyond the restricted regime
of static spherically-symmetric hairy black-hole spacetimes. To that
end, we have studied analytically the physical properties of the
recently discovered cloudy Kerr black-hole spacetimes. These are
{\it non}-static {\it non}-spherically symmetric Kerr black holes
\cite{Notec2} which support linearized massive scalar fields in
their exterior regions.

Using analytical techniques, we have established the fact that the
stationary bound-state massive scalar configurations which
characterize the rotating Kerr black-hole spacetime cannot be made
arbitrarily compact. In particular, we have derived the lower bound
[see Eq. (\ref{Eq34})]
\begin{equation}\label{Eq39}
{r_{\text{max}}\over{r_+}}>{{r_+}\over{r_-}}
\end{equation}
on the peak location $r_{\text{max}}$ of the radial eigenfunction
$\psi$ which characterizes the external bound-state massive scalar
configurations of the rotating Kerr black-hole spacetime.
Interestingly, the characteristic lower bound (\ref{Eq39}) is
universal in the sense that it is independent of the physical
parameters of the external massive scalar fields.

Furthermore, we have explicitly shown that the inequality
(\ref{Eq34}), which characterizes the composed
Kerr-black-hole-massive-scalar-field configurations, implies that
these non-static non-spherically symmetric \cite{Noteepm}
configurations respect the no-short hair lower bound (\ref{Eq1}).
Our results, together with the results presented in \cite{Hod11},
may therefore suggest that the lower bound
$r_{\text{hair}}>r_{\text{null}}$ may be a general property
\cite{Noteab} of asymptotically flat electrically neutral hairy
black-hole spacetimes.

It is important to emphasize the fact that the composed
Kerr-black-hole-massive-scalar-field configurations that we have
studied in the present paper respect the assumption made in the
original no-short hair theorem \cite{Hod11} that the energy density
$T^t_t$ outside the black-hole horizon approaches zero
asymptotically {\it faster} than $r^{-4}$. On the other hand, in the
charged (Kerr-Newman) black-hole case studied in \cite{Hodaa} the
electromagnetic energy density outside the black-hole horizon is
characterized by the asymptotic behavior $T^t_t\sim Q^2/r^4$, where
$Q$ is the electric charge of the black-hole spacetime. Thus,
charged Kerr-Newman black holes do {\it not} respect the assumption
made in the original no-short hair theorem \cite{Hod11} that the
energy density $T^t_t$ outside the black-hole horizon approaches
zero asymptotically {\it faster} than $r^{-4}$. The different
asymptotic behavior of the energy density in the charged case
studied in \cite{Hodaa} (as compared with the neutral case
considered in the present work) allows charged
Kerr-Newman-black-hole-charged-massive-scalar-field configurations
to violate the no-short hair bounds \cite{Hodaa}. It is therefore
important to emphasize that the results of the present paper are
restricted to the neutral (Kerr) black-hole case.

Finally, we would like to emphasize again that in this paper the
external scalar clouds (the stationary bound-state massive scalar
configurations) were treated at the linear level. As we explicitly
demonstrated, the main advantage of this approach lies in the fact
that the physical properties of the composed
Kerr-black-hole-linearized-massive-scalar-field configurations can
be studied {\it analytically}. We believe that, using {\it
numerical} techniques \cite{HerR}, it would be highly interesting to
further test the validity of the no-short hair lower bound
(\ref{Eq1}) in the non-linear regime of hairy
Kerr-massive-scalar-field black holes.

\newpage

\bigskip
\noindent
{\bf ACKNOWLEDGMENTS}
\bigskip

This research is supported by the Carmel Science Foundation. I thank
Yael Oren, Arbel M. Ongo, Ayelet B. Lata, and Alona B. Tea for
stimulating discussions.



\begin{thebibliography}{99}
\bibitem{Noteeln} We refer here to electrically
neutral asymptotically flat black-hole spacetimes.
\bibitem{Notereg} The term `regular black holes' is used in this paper
to describe black-hole spacetimes which are characterized by regular
event horizons.
\bibitem{Notevio} It is interesting to note that most of these
hairy black-hole solutions of the coupled Einstein-matter field
equations are known to be dynamically unstable.
\bibitem{Notec1} This requirement implies that the black-hole spacetime has
no extra globally conserved charges (besides the ADM mass) defined
at asymptotic infinity.
\bibitem{Notene} That is, rapidly-rotating black holes with the
property $J/M^2\simeq1$ (Here $J$ and $M$ are respectively the
angular momentum and the mass of the spinning black hole).
\bibitem{Noteal} It is worth emphasizing that our analytical proof (to be given below) would be
valid for {\it generic} values [that is, $J/M^2\in (0,1)$] of the
dimensionless black-hole rotation parameter.
\bibitem{Noteunits} We shall use natural units in which $G=c=\hbar=1$.
\bibitem{Notesmp} We shall assume $a>0$ and $m>0$ without loss of
generality.
\bibitem{Notedim} Note that the mass parameters $\mu$ of the scalar field stands
for $\mu/\hbar$. This field parameter therefore has the dimensions
of $($length$)^{-1}$.
\bibitem{Notehe} It is worth noting that the term `hair' is usually used in
the black-hole physical literature to describe the non-linear
external matter fields of a non-vacuum black-hole spacetime.
\bibitem{Noteanz} The integer parameters $m$ and $l\geq|m|$ are
the harmonic indices (azimuthal and spheroidal) of the scalar field
mode [see Eq. (\ref{Eq9}) below].
\bibitem{Notesa} These characteristic angular eigenfunctions are known as the spheroidal harmonic functions.
\bibitem{Notebj} We shall henceforth omit the angular harmonic indices $m$ and $l$ for brevity.
\bibitem{Notepp} One can assume $R(r=r_+)\geq0$ without loss of generality.
\bibitem{Notemap} Note that the transformation (\ref{Eq14}) maps the radial coordinate $r\in [0,\infty]$
into $y\in [-\infty,+\infty]$.
\bibitem{Notesi} Note that the lower bound (\ref{Eq20}) can be approached in
the eikonal $l=m\gg1$ regime, in which case one finds
$K_{mm}=m^2[1+O(m^{-1})]-a^2(\mu^2-\omega^2_c)$ \cite{Hodpp}.
\bibitem{Noteyas} Note that the near-horizon region (\ref{Eq23}) corresponds to $y\to
-\infty$ [see Eq. (\ref{Eq22})], which implies $e^{\tau y/r_+}\to
0$.
\bibitem{Noteab1} Here we have used Eqs. (9.1.54) and (9.6.3) of \cite{Abram}.
\bibitem{Notesk} The second mathematical solution of (\ref{Eq25}) is given by the modified
Bessel function of the second kind:
$\psi_2(y)=K_0(2\sqrt{{{F}/{\tau}}}e^{\tau y/2r_+})$. However, using
Eq. (9.6.13) of \cite{Abram}, one finds that this solution does not
respect the boundary condition (\ref{Eq12}) at the black-hole
horizon [In particular, one finds $\psi_2(y\to -\infty)\to -\tau
y/2r_+\to\infty$ at the black-hole horizon].
\bibitem{Noteep1} In particular, it is worth emphasizing again that (\ref{Eq28}) implies that
$\psi$ is a positive increasing function in the near-horizon
$x\ll\tau$ region.
\bibitem{Noteep2} In particular, it is worth emphasizing again that the stationary
bound-state configurations of the massive scalar fields are
characterized by exponentially decaying radial eigenfunctions at
spatial infinity [see Eq. (\ref{Eq11})].
\bibitem{Notebr} Note that ${{d^2\psi}/{dy^2}}>0$ in the
near-horizon $r\to r^+_+$ limit [see Eq. (\ref{Eq28})], whereas
${{d^2\psi}/{dy^2}}<0$ at the maximum point $r=r_{\text{max}}$.
Thus, the characteristic radial eigenfunction $\psi$ must have an
inflection point (with ${{d^2\psi}/{dy^2}}=0$) somewhere between the
horizon ($r=r_+$) and the maximum point ($r=r_{\text{max}}$).
\bibitem{Notels} It is worth emphasizing that this inequality is valid for all values of
the dimensionless black-hole rotation parameter $a/M$ [see Eq.
(\ref{Eq34})].
\bibitem{Noteab} That is, even beyond the restricted regime considered in \cite{Hod11} of static spherically-symmetric
hairy black-hole spacetimes.
\end{thebibliography}
\end{document}